\documentclass[12pt,preprint]{aastex}
\usepackage{graphicx}

\newcommand{\Msun}{M_{\odot}}

\shorttitle{Recurrent Nova Expansion}
\shortauthors{Toraskar et al.}

\begin{document}

\title{Dynamical Fragmentation of the T Pyxidis Nova Shell During Recurrent Eruptions \footnote{This paper is respectfully dedicated to the memory of Waltraut Seitter, co-discoverer of the ejecta of T Pyxidis.}}

\author{Jayashree Toraskar, Mordecai-Mark Mac Low, Michael M. Shara, David Zurek}
\affil{Department of Astrophysics, American Museum of Natural History,
  79th Street at Central Park West, New York, NY 10024-5192}
\email{toraskar@amnh.org, mordecai@amnh.org, mshara@amnh.org,dzurek@amnh.org}

\begin{abstract}

{\em Hubble Space Telescope} images of the ejecta surrounding the nova T Pyxidis resolve the emission into more than two thousand bright knots. We simulate the dynamical evolution of the ejecta from T Pyxidis during its multiple eruptions over the last 150 years using the adaptive mesh refinement capability of the gas dynamics code Ramses.  We demonstrate that the observed knots are the result of Richtmeyer-Meshkov gas dynamical instabilities (the equivalent of Rayleigh-Taylor instabilities in an accelerated medium).  These instabilities are caused by the overrunning of the ejecta from the classical nova of 1866 by fast moving ejecta from the subsequent six recurrent nova outbursts. The model correctly predicts the observed expansion and dimming of the T Pyx ejecta as well as the knotty morphology. The model also predicts that deeper, high resolution imagery will show filamentary structure connecting the knots.  We show reprocessed Hubble Space Telescope imagery that shows the first hints of such structure.
\end{abstract}

keywords{: T Pyx/Recurrent Novae}

\section{Introduction}

Recurrent novae (RNe) are cataclysmic variables that display multiple eruptions on a timescale of  decades \citep{war95}. The same mechanism powers both recurrent and classical novae: thermonuclear runaways in the hydrogen-rich envelopes of white dwarfs (WDs). The erupting envelopes have been accreted by the WDs from the non-degenerate companions---red dwarfs or red giants---in these binary systems. The observed short recurrence timescales of RNe are only possible for massive WDs ($ \sim 1.2$--$1.4~\mbox{ M}_\odot$) with high mass accretion rates ($\sim 10^{-8}~M_\odot$~yr$^{-1}$)) (\citet{sta85} and \citet{yar05}). Such high accretion rates suggest that RNe might be part of the population of Type Ia supernova progenitors \citep{kov94,dlv96}.  Ten RNe are known in the Milky Way; \citet{sch10} gives an extensive review of their observational properties.

The first RN discovered (and the prototype of such systems) was T Pyxidis (T Pyx), detected by H. Leavitt in 1913 \citep{pic13}. With six identical \citep{may67,sch10} recorded eruptions (1890, 1902, 1920, 1944, 1966 and 2011) to its credit, T Pyx remains the RN with the longest track record, and an average recurrence time of about 20 years. The puzzling lack of eruptions during the past 44 years, followed by its surprising 2011 eruption, have been explained by T Pyx's sharply decreasing luminosity (a factor of six) and mass accretion rate (a factor of 30) during the same time period \citep{sch10}. The secularly decreasing accretion rate implies, in turn, that T Pyx will not undergo another recurrent nova eruption \citep{sps10} before it drops into a state of hibernation (i.e. nearly zero mass accretion rate) lasting many millenia \citep{sha86}.

The ejecta of T Pyx, displaying a circular ring with a radius of 5'', were first detected by \citet{due79}. More sensitive observations have since revealed fainter outer material, reaching a radius of 6.5'' \citep{sps10}. This morphology was initially interpreted as an edge-brightened shell.  \citet{wil82} used spectroscopy of the ejecta to show that the material has roughly solar abundances and displays emission lines similar to those seen in planetary nebulae. \citet{sha89} demonstrated that the apparent shell was expanding too slowly to have originated in the 1944 or 1966 eruptions, and that it was at least twice as large as previously suspected. \citet{con97} showed that the spectral line fluxes can be successfully explained by shock heating when the fast ejecta from one nova eruption run into the slower ejecta from an earlier eruption. {\em Hubble Space Telescope (HST)} images from 1994 and 1995 resolved the previously observed ring into over 2000 knots, some of which fade or brighten significantly on a timescale of order one year \citep{sha97}. Expansion of the knots was not detected, though the time baseline used (1.7 yr) was quite short. 

Thirteen years later \citet{sps10} re-observed T Pyx with {\em HST}, detecting knot expansion velocities of 500 - 700 km/s . The observed fractional expansion of the knots is constant (hence there is little deceleration from the interstellar medium). The {\em HST} observations constrain the knots to have originated in an explosion close to the year 1866, and to possess a total mass of $\sim 10^{-4.5} \Msun$.  The distance to T Pyx is poorly constrained. \citet{sps10} estimate it to be $D = 3.5 \pm 1$~kpc. The 1866 event must have been a classical nova eruption, preceded by a low accretion epoch lasting of order 1 Myr \citep{sps10}. An 1866 nova event would have triggered a supersoft X-ray source on the T Pyx WD, driving a much higher mass transfer rate shortly after that event. Mass transfer has been steadily declining  \citep{sch05} in the T Pyx binary, and in 2009 was a factor of 30 less than in 1890. No more RN outbursts are expected before T Pyx enters a state of hibernation, expected to last for 2--3 Myr. There is consensus that the WD in T Pyx is very unlikely
to exceed the Chandrasekhar mass and to become a Type Ia supernova \citep{sel08} and \citet{sps10}).

One critical test of the scenario in which the ejecta are shaped by the history of a nova followed by six recurrent nova eruptions is to reproduce the morphology of the ejecta.  We hypothesize that the observed knots are produced by Richtmeyer-Meshkov instabilities excited in the thin, swept-up shell of circumstellar gas surrounding the nova ejecta when subsequent explosions accelerate it.  This instability is the equivalent of the Rayleigh-Taylor instability that occurs when acceleration rather than gravity drives overturn of a dense fluid supporting a more rarefied fluid \citep{ric60,mes69}.  Outward acceleration of the dense shell by more rarefied ejecta results in an effective gravity pointing inward, from the shell into the ejecta, and thus results in overturn and fragmentation of the dense shell.

We test our hypothesis using three-dimensional, gas dynamical simulations that include radiative cooling and are sufficiently well resolved numerically to follow the instabilities that can cause the observed fragmentation.  \citet{gar04} previously reported a preliminary two-dimensional model of this problem. In Section 2 we outline the simulation techniques, and describe the initial velocities, time intervals and ejected masses we assumed. These were taken from the observations of T Pyx by \citet{ada20}, \citet{joy45}, \citet{cat69} and \citet{sps10}. Our numerical results are presented and compared with the {\em HST} observations of T Pyx in Section 3. Our results and conclusions are summarized in Section 4.

\section{Simulations}

\subsection{Code Used}

We compute a fully three dimensional model of the original T Pyx classical nova of 1866, and the subsequent recurrent novae of 1890, 1902, 1920, 1944, and 1966 using a slightly modified version of the gas dynamical code RAMSES, version 3.0 \citep{tey02}. This code uses adaptive mesh refinement, with a tree-based structure that allows recursive grid refinement on a cell by cell basis. The hydrodynamical solver is based on the second order Godunov method. The conservative variables are taken to be piecewise constant over the mesh cells at each time step and the time solution is determined by the exact solution of the Riemann problem at the intercell boundaries (use of the exact Riemann solver is one of the choices in the default code).          

\subsection{Problem Setup} 

We modified the code to implement repeated instantaneous energy inputs to simulate nova explosions. A driver was added that set up nova explosions at given times, which were then allowed to evolve over time without any further changes until the next explosion. The nova explosions were located at the center of the cubical grid. The source was assumed to have finite size in order for the AMR code to be able to resolve its evolution. The density and the velocity of the source material were derived from the observed mass and velocity of the ejecta as we explain below.

The nova explosion was assumed to be instantaneous, and after the explosion, the system developed naturally. The densities calculated as explained above were used for actual expansion studies. The additional parameters added to the Ramses input file were source density, source velocity, and time step for the eruption.  

A cubical grid with an edge length of $6.0 \times 10^{17}$~cm was used for the simulation.  Our most refined level had resolution equivalent to 512$^3$ zones on a single grid. The nova source was located at the center of the cube. It had a radius $r_s = 1.25\times 10^{16}$~cm, which is about 11 pixels on our most refined level.  

To determine the background density we determined the density that would be consistent with the size of the observed shell and the energy of the 1866 nova explosion.  Inverting the \citet{sed59} similarity solution for the radius of a blast wave to solve for the density,
\begin{equation}
\rho = 2.2  E { t^2 }/ {r^5}
\end{equation}
where $r$ is the radius of the expanding shell,$ E = 7.46 \times 10^{43}$~g~cm$^{2}$~s$^{-2}$ is the kinetic energy of the first nova, and the age of the shell $t = 129$~yr.
We take the distance to T Pyx to be $4.5$ kpc \citep{sho11}, although this is a value uncertain to a factor of two.  If we then take the angular radius of the shell to be 6.5''  \citep{sps10} we can use Equation~\citep{sed59} to derive a background density $\rho_0 =1.689 \times 10^{-25}$~g~cm$^{-3}$.
The background gas was set to be uniform, with a pressure of  $1.3806 \times 10^{-12}$~erg~cm$^{-3}$, yielding a temperature of $1.0 \times 10^5$~K.  This rather high temperature was set by accident, but does not appear to markedly influence the shell dynamics of the nova that we study.  It could certainly be consistent with a location of the nova within a background superbubble.  Finally, we set up random zone-to-zone noise at a level of $30$ \% of the background density for the 1866 nova and $50$ \% for all other novae in order to trigger the shell instabilities. 

We assumed the gas to be monatomic, with adiabatic index $\gamma = 5/3$. We used the equilibrium ionization cooling implemented in Ramses, which uses the lookup table described by \citet{cor04}.  To balance the cooling in the background gas, we also use the photoionization heating implemented in Ramses, parameterized by  a background UV flux with power-law spectrum and normalization J21.  We determined empirically that $J21 = 1$ provided approximate thermodynamic balance, and used that value in our runs. 
            
The density of the explosion source $\rho_s$ was found by setting the value of the kinetic energy $E$ and the expansion velocity $u_0$ of each nova outburst, so that
\begin{equation}
  \rho_s=\left(\frac{2 E (\Delta x)^3}{ \Sigma v_{ijk}^2}\right),
 \end{equation} 
where $\Delta x$ is the length of  a single zone, $v_{i,j,k}$ is the velocity in the zone with index $(i, j, k)$, and the sum runs over all the zones in the source and transition regions.  The kinetic energy $E$ is derived from the mass and velocity measured by the observations. 

For the 1866 nova the likely ejecta mass was $\sim 3 \times 10^{-5}~M_\odot$, and velocity was $u_0 = 5 \times 10^2$~km~s$^{-1}$ \citep{sps10}. All the recurrent novae ejecta were assumed to have  masses of  $\sim 1 \times10^{-7}~M_\odot$, and ejection velocities of $2 \times 10^3$~km/sec. Our simulation stops just before the 2011 eruption, whose effects on the ejecta will not become significant for several decades.

The T Pyx ejecta are observed to be elongated in the polar directions \citep{sha97}. We assume that this asymmetry is caused by an asymmetric nova explosion, so we implant it directly in the source velocity field for the first eruption in 1866, so that we set
\begin{equation}
u_1 (\phi) =            u_0(1.0  + 0.1 \cos \phi).
\end{equation}
The subsequent explosions were assumed to be spherical.  We initialize each zone in the source region with the Cartesian components of the derived velocity $u_1$ or, for the subsequent explosions, $u_0$.

The source region was extended by 50\%  as a transition region between the high velocity material at the edge of the source region and the material at rest outside.  We found this necessary to maintain the stability of the Riemann solver. To do this,
we multiplied  velocities by a radial profile function, so that $v_{ijk} = u_{0,1} \xi(r)$ as appropriate.  We chose the profile to dip to zero at the center of the source to avoid Riemann solver failures caused by a strong rarefaction wave at the center of the source region, and have a smooth transition to zero across the outer transition region.  The actual form of the profile function is  
\begin{equation}
\xi(r) = \left\{ 
       \begin{array}{ll} 
             (\tanh [10 (r/r_s - 0.5) ]+ 1.0)/2, & \mbox{ if } r < r_s \\
             (\tanh[ 10 (r/r_s - 0.5) ] + 1.0) *(1.5r_s/r-1) ,  & \mbox{ if }r_s < r < 1.5 r_s \\
             0.0,                                                       & \mbox{ if } r > 1.5 r_s.
          \end{array}  
          \right.
\end{equation}
The density within the transition region was set to the source density $\rho_s$.

\section{Observations} 

\begin{figure}
\epsscale{1.0}
\plotone{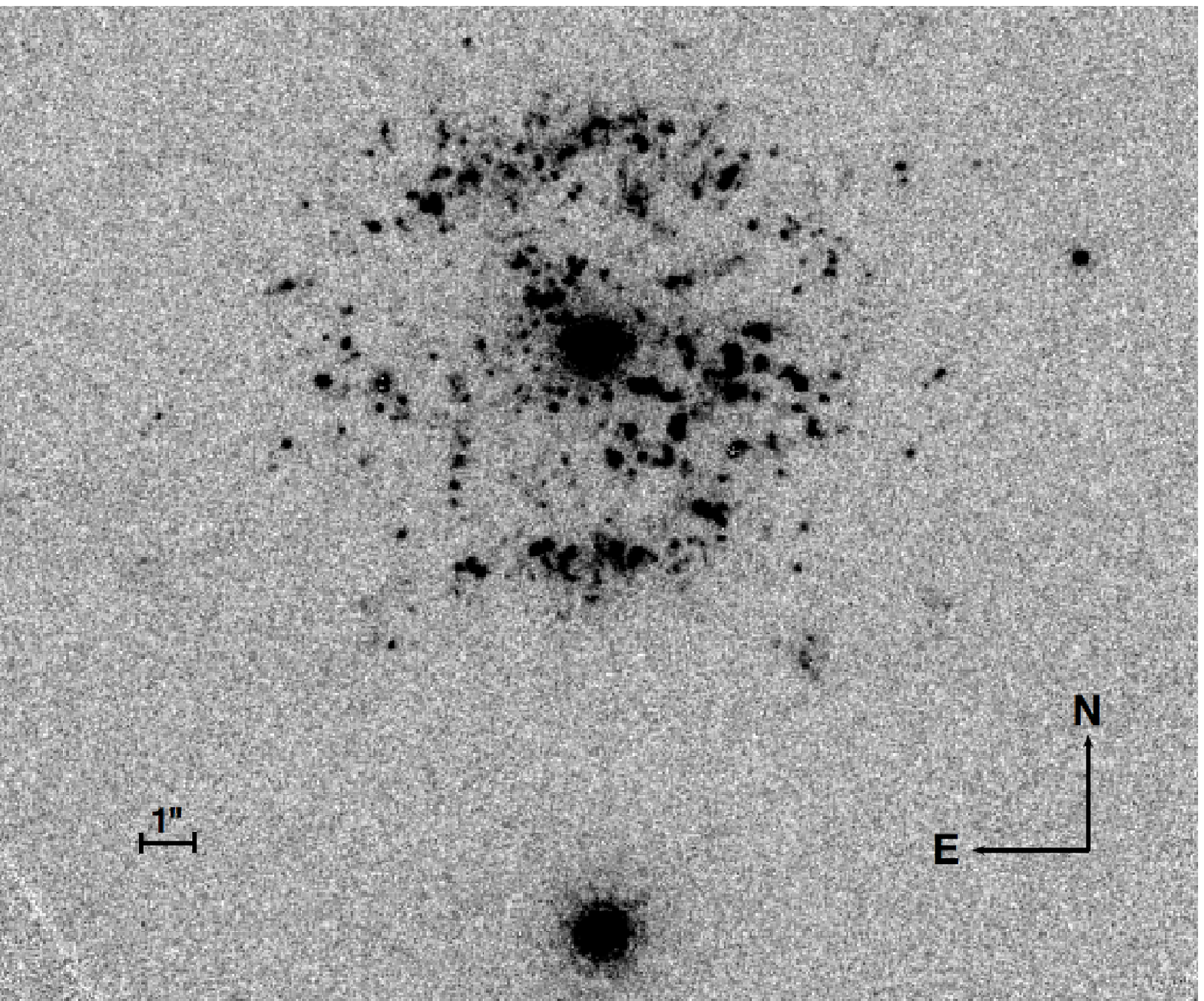}
\caption{A drizzled image of the combined 1994 and 1995 {\em HST} WFPC2 [NII] imaging of T Pyx.
\label{HST [NII] T Pyx}
}
\end{figure}

We extracted the deepest set of {\em HST} images of T Pyx, taken with the Wide Field and Planetary Camera2 through the F658N filter, which isolates the strongest optical [NII] emission line. Details of the images are given in Table 2 of \citep{sha97}. These were drizzled \citep{fru02} to produce Figure~\ref{HST [NII] T Pyx}. Figure 1 reveals faint, emitting blobs as much as 0.5 mag fainter than the faintest blobs seen in the (undrizzled) F658N image that is shown in Figure 2 of \citet{sha97}. 

 \section{Comparison}

\begin{figure}
\epsscale{.70}
\plotone{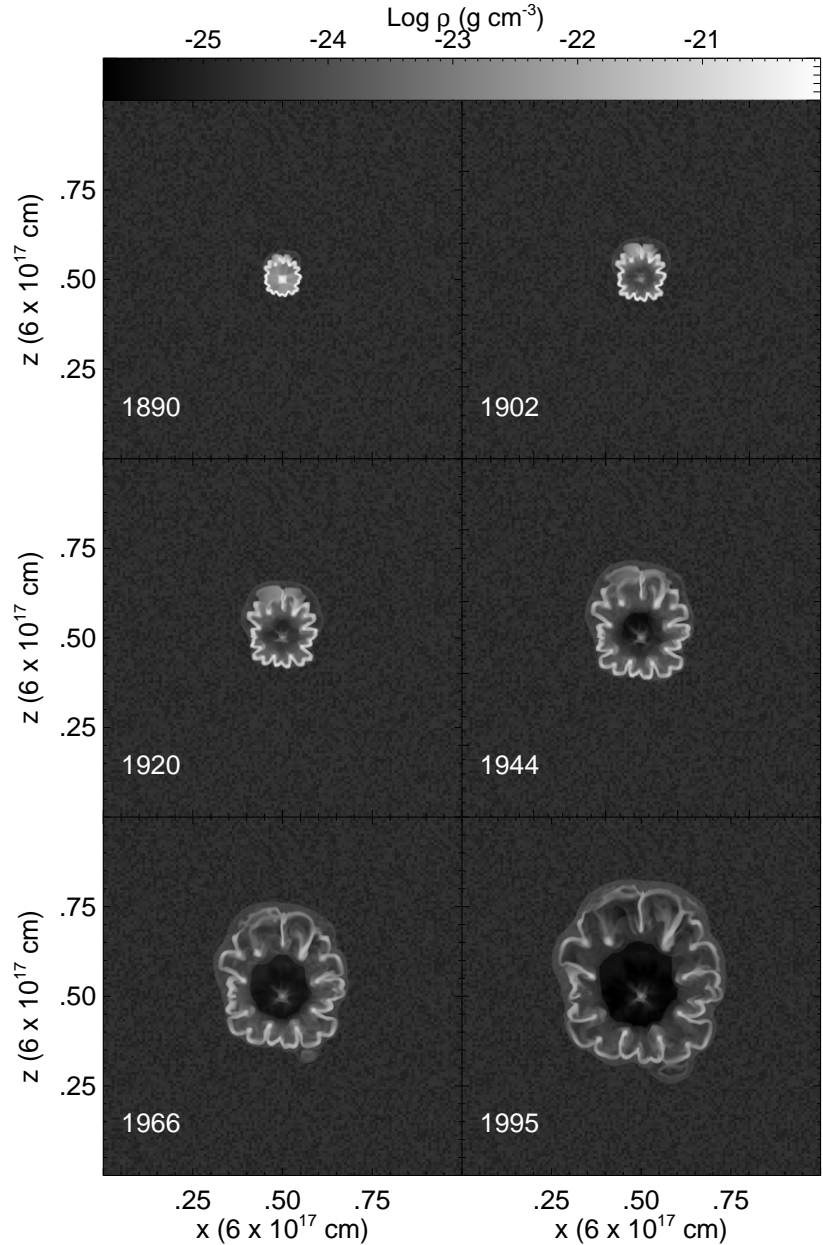}
\caption{Log density distribution for the six epochs just before the next nova is about to explode, except for the final epoch which is 1995. Shown are two-dimensional cuts through the log density field
$xz$-plane, showing the imposed asymmetry.  The color scale shows log of density with values given by the color bar. The figure demonstrates the fragmentation of the nova shell by instabilities as it is repeatedly accelerated. 
\label{six_novae-zcuts}
}
\end{figure}

The ejecta from each of the eruptions described above was allowed to expand in the simulations for 141 years after the first (classical nova) eruption, reaching the year 2007, to enable a comparison with the {\em HST} observations. 
Figure~\ref{six_novae-zcuts} displays polar cuts through the computed log density distribution of the ejecta of the first six eruptions, seen just before the next recurrent nova eruption, except for the last frame, which shows the distribution in the year 2007. The Figure shows the model prediction that the knots expand away from the center as time progresses, as recently demonstrated observationally by \citet{sps10}. The outer shell (i.e. the outermost part of the ejecta)  appears to be the remnants of the original 1866 nova eruption, moving slowly away from the central star. Our previous impression of many knots \citep{sha97} is reinforced.  The fingers that are so prominent in the simulation visualization of Figure 2 are not seen in Figure 1. We predict that deeper images of the T Pyx ejecta will show these elongated features.

\begin{figure}
\epsscale{.80}
\centering
\vspace{-2.5in}
\plotone{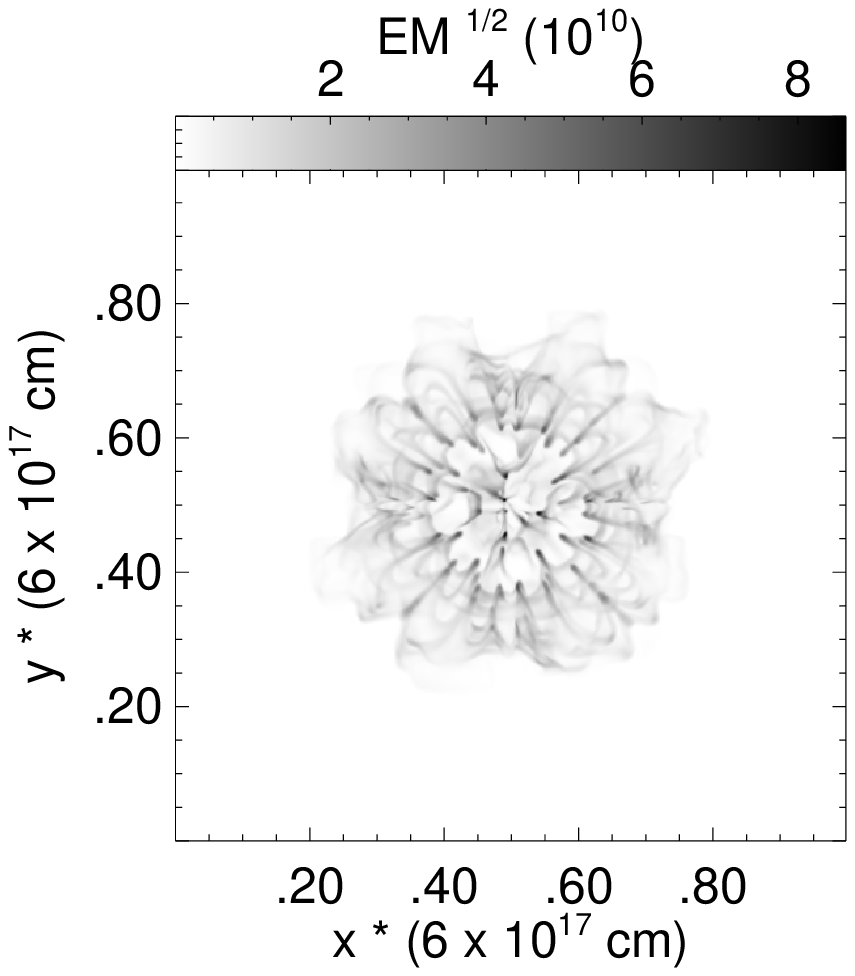} 
\vspace{-4in}
\plotone{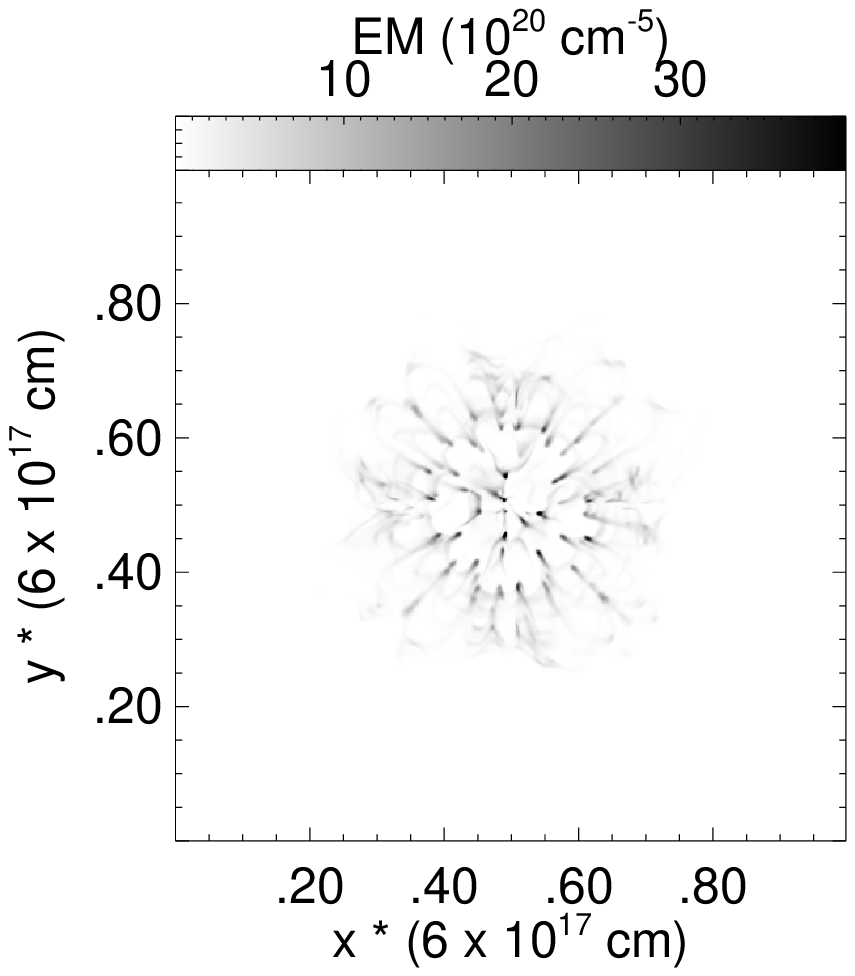}
\vspace{-1.5in}
\caption{(a) Square root and (b) linear display of the emission measure of the ejecta in 2007, as viewed along the direction of observation (the z axis of the simulation). The T Pyx source is at the center of the box. The source size is $2.5 \times 10^{16}$~cm. The box size is $6.0 \times 10^{17}$~cm. The color scales at the top indicate the range of values in cgs units.
\label{sqrt_emi3_z}
}
\end{figure}

\begin{figure}
\epsscale{0.70}
\plotone{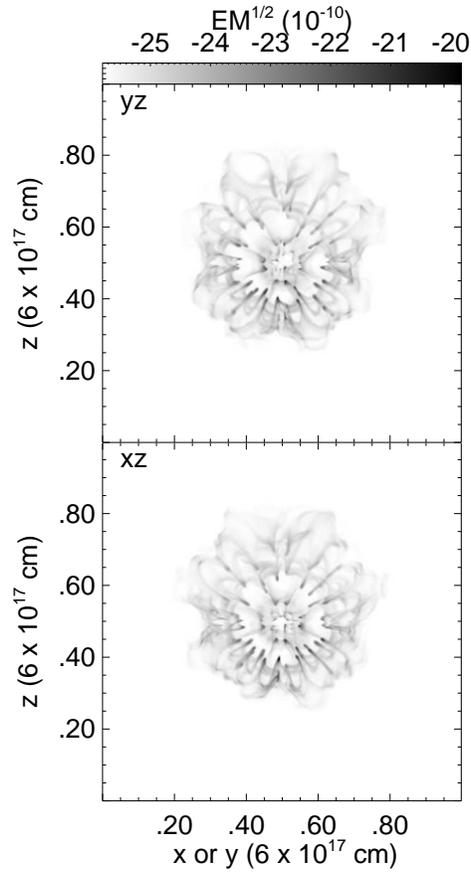}
\caption{Square root of the emission measure  for year 2007, as viewed along the x and y axes (perpendicular the the line of observation). The shapes of these distributions are very similar to the one shown in Figure~\ref{sqrt_emi3_z}a. This indicates that the observed results are not artifacts, and are not sensitive to the displayed line of sight. The color scales at the top indicate the range of values in cgs units.
\label{sqrt_emission_measure}
}
\end{figure}

Figures~\ref{sqrt_emi3_z} and~\ref{sqrt_emission_measure} display images from the year 2007 model of the square root of the simulated emission measure $L = \Sigma {\lambda \times n_{ijk}^2}$ along all three axes, where $n$ is the number density, $\lambda = {6.}^{17} / {512}$ gives the length of the unit cell. These were produced by interpolating the adaptive grid onto a uniform grid at the finest level, and then summing. The same filamentary structures seen in the density distributions of Figure 2 are clearly visible in these emission measure images.  They are here emphasized in Figure~\ref{sqrt_emi3_z}a by the square root scaling adopted.
Figure 3b shows a clipped display of the emission measure to again emphasize the weaker emission.  In this case we take the top of the grey scale to be ${4} \times 10^{21}$~cm$^{-6}$.

Figure~\ref{sqrt_emission_measure} shows the square root of the emission for year 2007, as viewed along the $x$ and $y$ axes (perpendicular to line of observation). 
Comparison of Figures~\ref{sqrt_emi3_z}a and~\ref{sqrt_emission_measure} suggests that the overall shape of the ejecta distribution is neither an artifact of, nor sensitive to the direction of observation despite the asymmetry of expansion velocity observed.

\begin{figure}
\epsscale{0.70}
\plotone{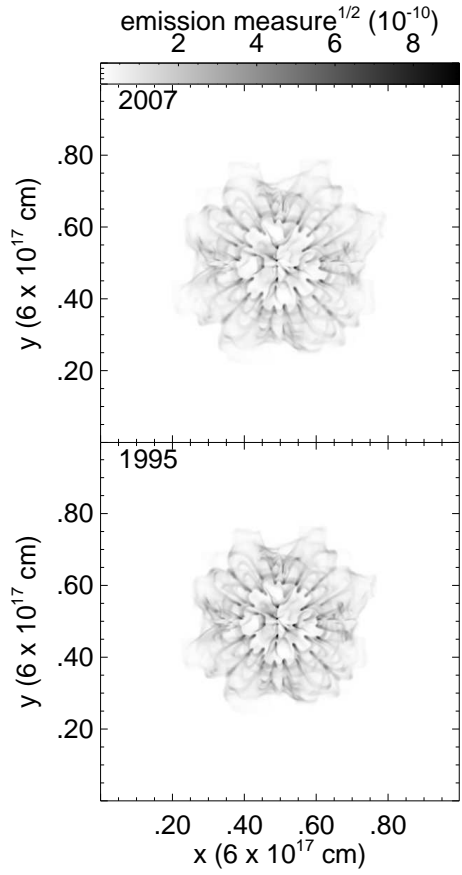}
\caption{Comparison of the $xy$-plane distributions of the square root of the emission measure  at years 1995 (bottom figure) and 2007 (top figure). The color scale at the top  indicates the square root  of the emission measure values. The color scales at the top indicate the range of values in cgs units.
 \label{EM_1995_2007}
 }
\end{figure}

Our model allows us to predict the changes in the nebula over the time span represented by the observations. In Figure~\ref{EM_1995_2007} we show the predicted behavior of the square root of emission measure between 1995 and 2007.  Slow expansion of the ejecta is predicted, while the overall shape remains very similar. The only {\em HST} observations of T Pyx available are for the epochs 1994, 1995 and 2007; using these data to directly compare the positions of selected knots, \citet{sps10} were able to demonstrate the knots' expansion. We show below that our models succeed in demonstrating the same expansion. 

Aside from morphology, another comparison that we can make between the model and the observations is the radial profile of azimuthally averaged surface brightness.  This comparison is complicated because the distance to T Pyx is only known to within a factor of two, so neither the physical scale nor the conversion between luminosity and observed flux can be well determined. In Figure~ 9, we assumed a distance of 4.5 kpc (as assumed in the background density calculation above) in order to set the physical scale.  As the conversion between H$\alpha$ surface brightness in the simulated observations and the actual [N II] surface brightness is also uncertain, we then simply scaled the model profile vertically to fit the intensity at the first peak of the observed profile in our models, to allow comparison of the measured, azimuthally-averaged radial surface brightness distribution observed in 1995 with the model prediction. There is general, overall agreement in the trend and shapes of the curves, but the full complexity of the observations is not reproduced by the simulation, which has somewhat fewer bumps and wiggles. Figure~\ref{Radial-mods}  is a prediction (based on Figure 3) of the azimuthally-averaged radial surface brightness profile of T Pyx expected in 1995, 2007 and 2011. The simulation predicts both a dimming and an expansion of the ejecta, in good agreement with the observations \citep{sps10}.

\section{Discussion}

Thousands of knots were found in the T Pyx ejecta by \citet{sha97}, while only around a hundred are seen in Figure 3.  This can be shown to be due to the limited numerical resolution we used in our simulation.  The Richtmeyer-Meshkov instability has no intrinsic minimum wavelength, similarly to the Rayleigh-Taylor instability. The only mechanism that limits fragmentation is viscosity. As a result, increasingly fine numerical grids, with decreasingly low numerical viscosity resolve increasingly fine wavelengths. We demonstrate this effect by comparing in Figure~\ref{resolution-study} our standard model with a model run at half resolution, having a finest grid resolution equivalent to $256^3$ zones but otherwise identical conditions.  The lower resolution model shows development of the instability, but with clearly decreased shell fragmentation because of the greater numerical viscosity acting on the evolution. Conversely, a greater resolution model will yield more fragmentation, and thus more knots.  (A dramatic example of this process in a different astrophysical context is shown in Figure~7 of \citet{fuj09}.)  This will provide a better match not only in the number of knots seen in Figure~1, but also to the radial profiles shown in Figures~7 and~8.  However, doubling the linear resolution would require a factor of 16 more computer time, requiring close to a million CPU-hours, a project beyond the scope of the present paper.

If the scale at which the physical viscosity dominates can be resolved, then its inclusion will limit this process to the correct physical result. The effective resolution of our $512^3$ zone simulation is $1.2 \times 10^{15}$~cm.  The physical viscosity will dominate when the Reynolds number Re$ = LV / \eta \sim 1$, where $\eta$ is the kinematic viscosity, and $L$ and $V$ are characteristic length and velocity scales. For dilute gas, the kinematic viscosity is \citep[e.g.]{zwi41} \begin{equation} 
\eta \sim (2 \times 10^{17} \mbox{cm}^{2} \mbox{ s}^{-1}) 
\left(\frac{T}{300 \mbox{ K}}\right)^{1/2}
\left(\frac{\rho}{10^{-21} \mbox{ g cm}^{-3}}\right),
\end{equation}
where we have used typical values for the shell density and temperature for scaling. The requirement that Re$\sim 1$ then gives a length scale of 
\begin{equation}
L \sim \eta / V = (2 \times 10^{12} \mbox{ cm} 
\left(\frac{T}{300 \mbox{ K}}\right)^{1/2}
\left(\frac{\rho}{10^{-21} \mbox{ g cm}^{-3}}\right)
\left(\frac{V}{1 \mbox{ km s}^{-1}}\right)^{-1},
\end{equation}
where we have roughly scaled the velocity by the sound speed in the cold gas.
This is clearly far shorter than we can numerically resolve, and indeed far shorter than the scale at which the knots are physically observed, which \citet{sps10} suggests is about 0.2'', or $10^{16}$~cm if we assume a distance to T Pyx of $D = 3.5$~kpc

This suggests that some other process is actually limiting fragmentation, with the most obvious candidate being magnetic fields in the swept up shell. \citet{cha61} uses a linear analysis to demonstrate that a magnetic field suppresses growth of the magnetized Rayleigh-Taylor instability at scales 
\begin{equation}
L < \frac{B^2}{g(\rho_1 - \rho_0)},
\end{equation}
where $B$ is the field strength, $g$ is the gravitational acceleration, and $\rho_0$ and $\rho_1$ are the mass densities of the light and heavy fluids. The nonlinear development of the magnetized Rayleigh-Taylor instability has been studied, for example, by \citet{sto07}.  Magnetized Richtmyer-Meshkov instability would be expected to behave essentially identically.  We can check this explanation for plausibility by making order of magnitude estimates of the fragmentation scale, shell density, and  acceleration.  As above, we take $L = 10^{16}$~cm, while the acceleration acts over less than a month, so that we can estimate $g= (100 \mbox{ km s}^{-1})/(10^6 \mbox{ s}) = 0.1 \mbox{ cm s}^{-2}$.  The shell density reaches at least $10^{-21} \mbox{g cm}^{-3}$ in our models, in the absence of magnetic field, although the field, if present, will act to limit its compression, so this is not a fully self-consistent estimate.  The density of the rarefied gas can be neglected for our purposes. These estimates would then yield a required magnetic field strength of
\begin{equation}
B \sim (L g \rho_1)^{1/2} = (1 \mbox{ mG}) 
\left(\frac{L}{10^{16} \mbox{ cm}}\right)
\left(\frac{g}{0.1 \mbox{ cm s}^{-2}}\right)
\left(\frac{\rho_1}{10^{-21} \mbox{ g cm}^{-3}}\right).
\end{equation}
This is easily consistent with amplification by compression within the shell of swept-up interstellar magnetic field from the ambient value of $3-5 \mu$G, given that the density has increased by well over two orders of magnitude from the ambient medium to the cold, compressed shell.  This possibility bears further investigation.

\begin{figure}
\epsscale{0.50}
\plotone{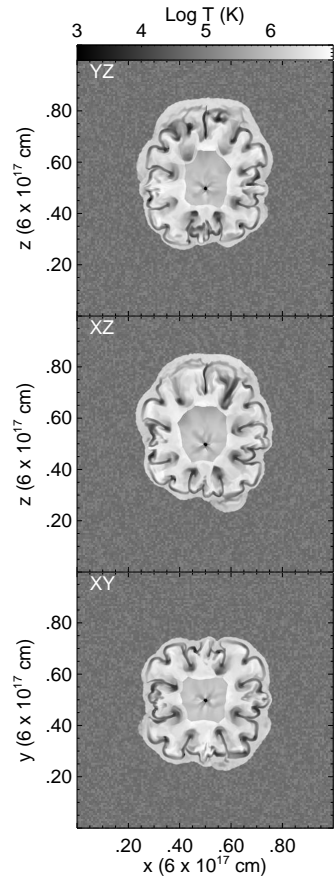}
\caption{Temperature distribution at the end of year 2007. Shown are the $xy$-plane,  $xz$-plane and $yz$-plane distributions. The color scale at the top indicates temperature values.
\label{temperature-cuts}
}
\end{figure}

The dense gas forming the knots originates in the shell swept up by the nova ejecta.  While the ejecta sweep up the shell, a strong reverse shock travels back through the much lower density ejecta.  The low densities result in inefficient radiative cooling, so the ejecta are heated to temperatures well over $10^6$~K, as can be seen in Figure~\ref{temperature-cuts}, which shows the temperature distribution of the ejecta at the end of 2007 in cuts across all three axes. While the temperature of the ejecta gas cannot yet be directly measured or compared with this figure, Figure 6 demonstrates that the visible knots should be surrounded by high temperature gas.  Their interaction should produce X-ray emission, particularly on the surfaces of the knots where the densities will be highest due to conductive evaporation \citep{cow77}.

\begin{figure}
\epsscale{0.95}
\vspace{-2in}
\plotone{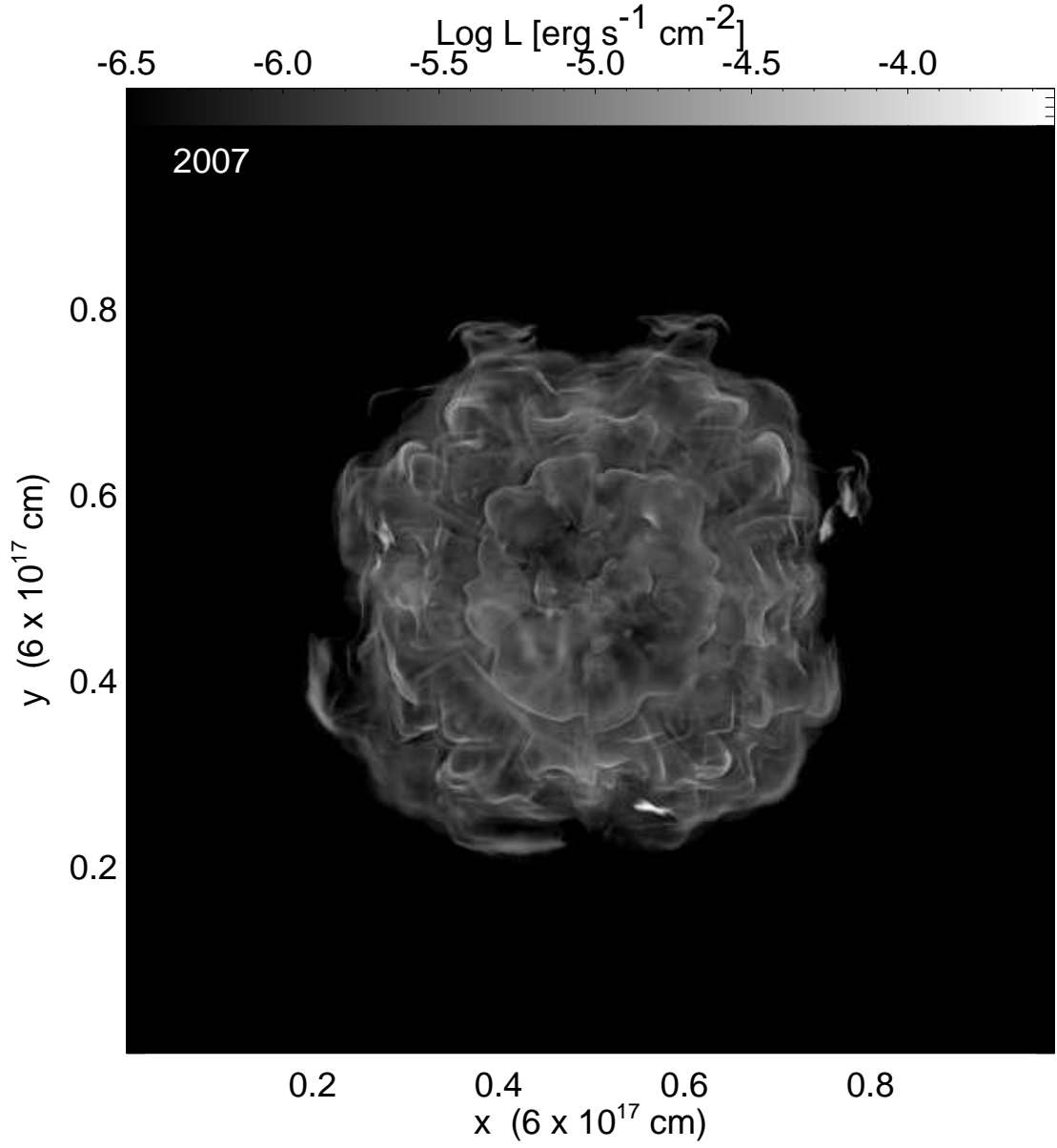}
\caption{Predicted X-ray flux in erg cm$^{-2}$ s$^{-1}$ at the end of year 2007. Shown is the line-of-sight  ($xy$-plane) X-ray flux distribution at the source. The color scale at the top indicates flux values.
\label{luminosity-cut}
}
\end{figure}

\begin{figure}
\epsscale{0.50}
\plotone{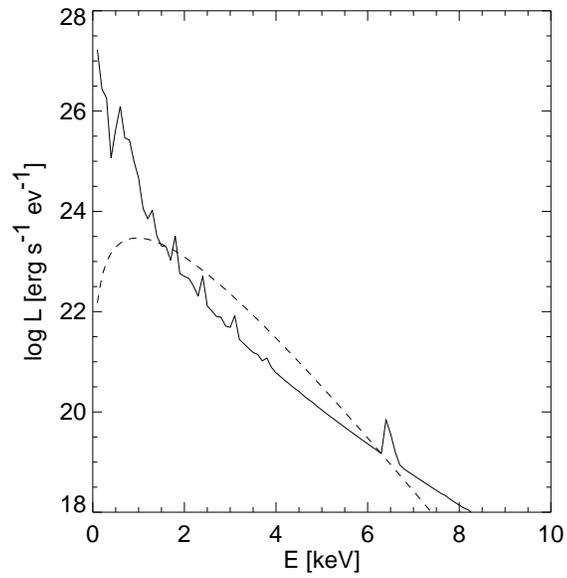}
\caption{Predicted X-ray luminosity spectrum in units of erg s$^{-1}$ ev$^{-1}$ calculated from the model temperature distribution using the \citet{ray77} code for the soft X-ray spectrum of a hot plasma.  A blackbody spectrum with a temperature of $T = 4\times 10^6$~K and an area of $10^3$ zones (dashed) is overlaid for comparison, demonstrating that a single temperature fit will not capture the behavior of this spectrum.
\label{X-ray_spec}
}
\end{figure}

    
Figure~\ref{luminosity-cut} displays the the X-ray luminosity along the line of sight. To calculate this we used the the October 1993 version of the \citet{ray77} code, with the default set of cosmic abundances from \citet{allen73}. We calculated the emission over the energy range 100 eV to 10 KeV.  We included no absorption from neutral hydrogen, no enrichment of metals in ejecta, and no contribution from the central object or from photo ionization. The total luminosity over this waveband at the end of 2007 was calculated is $2.48 \times 10^{29}$ erg~s$^{-1}$. We note that a great deal of the extended luminosity actually comes from a small number of hotspots.  Nevertheless, the total luminosity is much less than the value found by \citet{bal10} of $6 \times 10^{32}$ erg~s$^{-1}$, suggesting that the emission they report comes from the central object, not from extended emission from the ejecta. \citet{sokoloski12} reports only emission from the central object in the energy range 0.5--8~keV. We find an order of magnitude less emission in this harder waveband, because of the relatively low effective emission temperature.  They do not give an explicit lower limit to the extended emission they can measure, but it is likely higher than our predicted value.  We also use the same technique to predict the X-ray spectrum over this waveband at the same time, as shown in 
Figure~\ref{X-ray_spec}. 

We compare this spectrum to a blackbody spectrum to estimate how well or poorly a single temperature model reproduces our spectrum.  In Figure~\ref{X-ray_spec} we overplot a blackbody spectrum with a temperature of $5\times 10^6$~K, a value chosen by eye to fit the slope at energies of 2--6 keV.  The blackbody spectrum is normalized by an estimate of the area of the emitting zones, which is rather arbitrary, so the shape is the actual point of comparison rather than the absolute normalization.  The comparison reveals that the actual non-isothermal temperature distribution shown in Figure~\ref{temperature-cuts} produces a substantial soft X-ray excess compared to a single-temperature model. 


\section{Summary } \label{summary}

We have carried out fully three-dimensional, gas dynamical simulations of the evolution of the ejecta of T Pyx, starting with its classical nova outburst in 1866 and continuing up to the time of the latest {\em HST} observations in 2007. We are able to predict not only the observed expansion of the ejecta, but also the extremely knotty morphology. Our simulations demonstrate that the knots form when ejecta from the later outbursts collide with the swept-up, cold, dense shell from the nova explosion and drive Richtmyer-Meshkov instabilities in it. The resulting overturn and fragmentation leaves dense knots connected by filamentary structures and surrounded by hot gas. The fragmentation scale observed can best be explained by a swept-up interstellar magnetic field limiting fragmentation at the smallest scales.  We predict that deeper optical observations of T Pyx's ejecta using {\em HST} or other instruments will show filamentary structures connecting the knots, while X-ray observations will be able to find evidence of the interaction between the hot gas and the dense cold knots.  

\acknowledgments JT, DZ and MS were partly supported by HST grant HST-GO-12446.  M-MML was partly supported by NSF grant AST11-09395. Computations were performed on an Ultrasparc III cluster generously donated by Sun Microsystems.

\begin{figure}
\epsscale{1.0}
\plotone{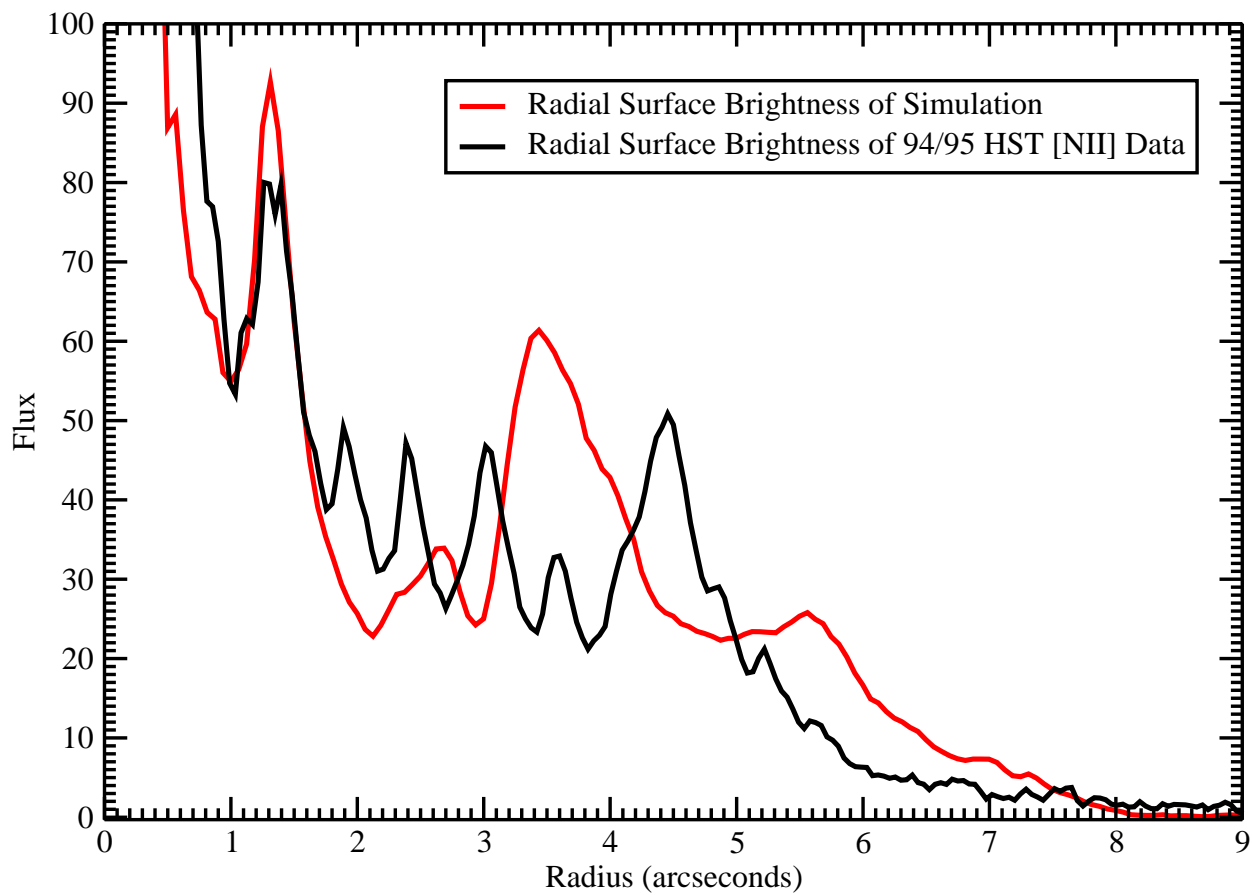}
\caption{The observed, azimuthally-averaged, radial surface brightness distribution of the extended nebulosity surrounding T Pyx (black line) and the same quantity from our simulation at 1994/1995 (red line). The flux of the simulation has been divided by $10^{19}$ for comparison purposes and the radial scale has been adjusted (to account for uncertainties in the distance to T Pyx) to line up the feature at about 1.4 arc seconds from the central star.
\label{Radial-lum}
}
\end{figure}

\begin{figure}
\epsscale{1.0}
\plotone{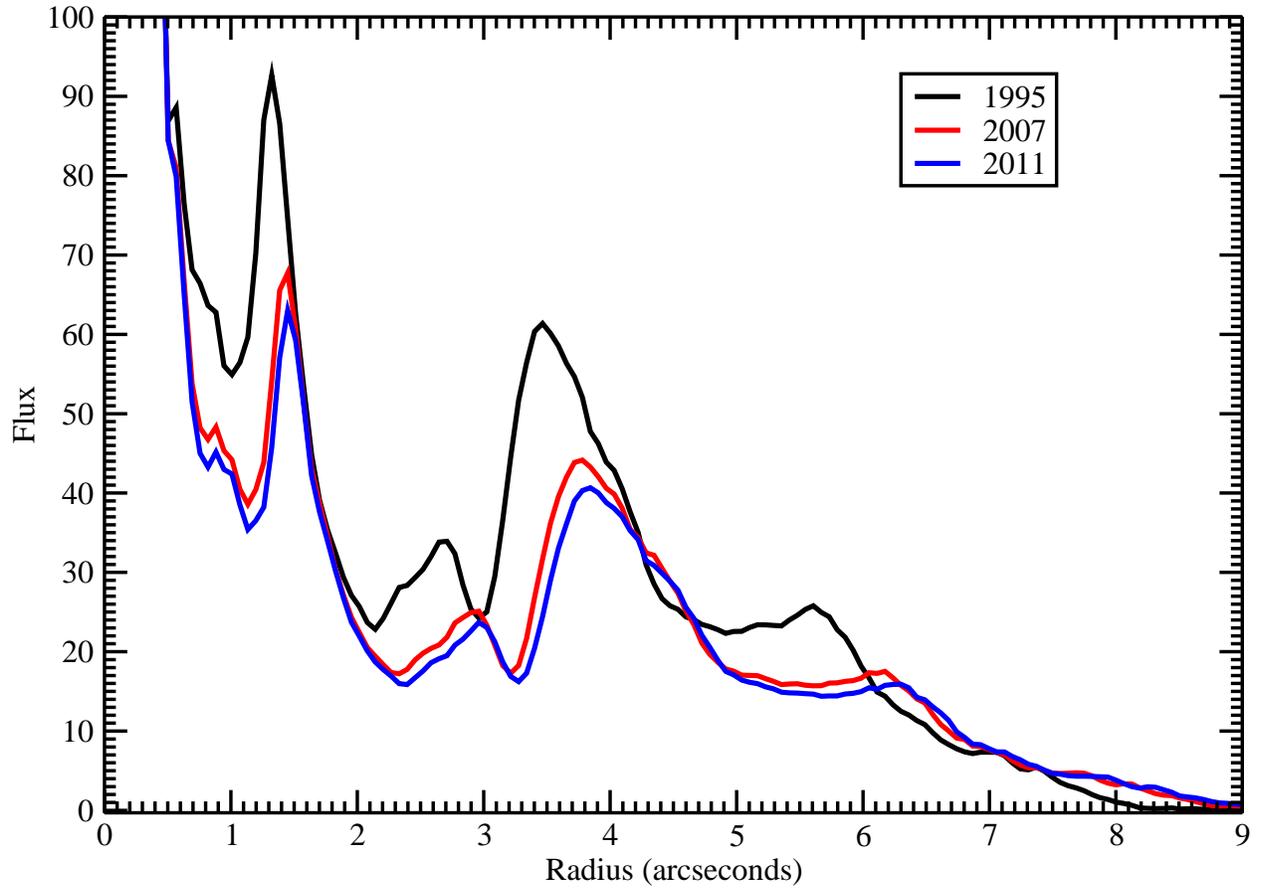}
\caption{The radial surface brightness of the nebulosity of the simulations corresponding to 1995, 2007 and 2011. The flux has been divided by $10^{19}$ and the x axis has been scaled as in figure 8. The key result is that both the expansion and the dimming of the T Pyx ejecta are clearly seen.
\label{Radial-mods}
}
\end{figure}

\begin{figure}
\epsscale{0.65}
\center
\plotone{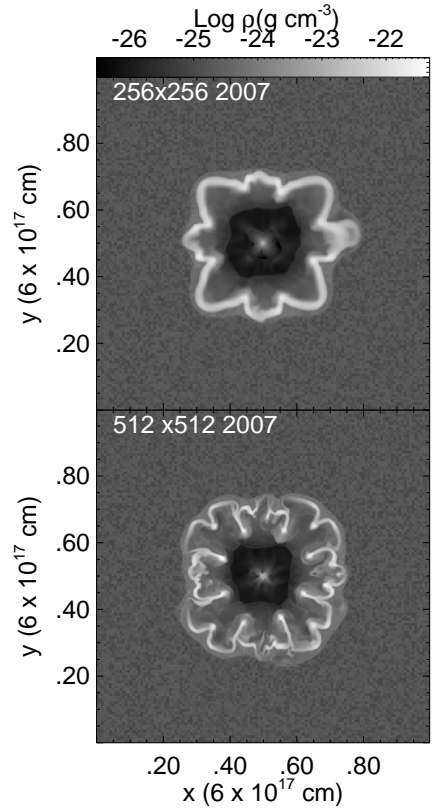}
\caption{Resolution Study on year 2007 log densities for $256^3$ and $512^3$ grids.
Shown are cuts along  the  $xy$-plane of log density distributions.
The color scale at the top indicates density values.  This shows the effect of increasing numerical resolution on instability wavelength and resulting fragmentation.
\label{resolution-study}
}
\end{figure}

\end{document}